\documentclass[pra,twocolumn,showpacs]{revtex4}

\usepackage{amsmath,amssymb,amsthm,epsfig,graphicx}
\usepackage[latin1]{inputenc}

\usepackage{xspace}
\usepackage{latexsym}
\usepackage{amstext}
\usepackage{amsxtra}

\newcommand{\Z}{\mathbb{Z}}
\newcommand{\R}{\mathbb{R}}

\newcommand{\C}{\mathbb{C}}

\newcommand{\F}{\mathcal{F}}
\newcommand{\E}{\mathcal{E}}
\newcommand{\ep}{\varepsilon}

\def\XXint#1#2#3{{\setbox0=\hbox{$#1{#2#3}{\int}$}
     \vcenter{\hbox{$#2#3$}}\kern-.5\wd0}}

\numberwithin{equation}{section}

\begin{document}
\bibliographystyle{unsrt} 

\title{Lowest Landau Level vortex structure of a Bose-Einstein condensate rotating in a harmonic plus quartic trap}
\author{Xavier Blanc, Nicolas Rougerie}
\affiliation{Universit{\'e} Paris 6, Laboratoire Jacques-Louis Lions,\\ 175 rue du Chevaleret, 75013 Paris, France. }

\date{\today}

\begin{abstract}
We investigate the vortex patterns appearing in a two-dimensional annular Bose-Einstein condensate rotating in a quadratic plus quartic confining potential. We show that in the limit of small anharmonicity the Gross-Pitaevskii energy can be minimized amongst the Lowest Landau Level wave functions and use this particular form to get theoretical results in the spirit of [A. Aftalion X. Blanc F. Nier, Phys. Rev. A 73, 011601(R) (2006)]. In particular, we show that the vortex pattern is infinite but not uniform. We also compute numerically the complete vortex structure: it is an Abrikosov lattice strongly distorted near the edges of the condensate with multiply quantized vortices appearing at the center of the trap.
\end{abstract}


\maketitle

\section{Introduction}\label{sec:Intro}

When rotated, superfluids are expected to exhibit a rich structure of quantized vortex states. This has been experimentally confirmed for Bose-Einstein condensates: the techniques developed to create a single vortex state \cite{MAHHWC,MCWD} rapidly led to arrays containing up to several hundreds of vortices \cite{ARVK,RAVXK,HCEC,ECHC}. Typically these condensates are confined by harmonic potentials, and the large vortex arrays are obtained for angular velocities $\Omega$ approaching the radial trap oscillator frequency $\omega$.\\ 
With these type of potentials, the limit $\Omega \rightarrow \omega $ is singular: as the centrifugal force balances the trapping force the Thomas-Fermi radius of the condensate diverges and the central density decreases towards zero.\\
In the experiments \cite{Exp1,Exp2}, a blue detuned laser directed along the axial direction adds a quartic component to the usual harmonic potential, resulting in a potential of the form :
\begin{equation}\label{eq:potentiel}
V(r)=\frac{1}{2}m\omega^2 r^2 + \frac{k}{4}r^4
\end{equation}
where $m$ is the mass of the atoms. The nice feature of this potential is that the centrifugal force, varying as $\Omega^2 r$ can always be compensated by the trapping force varying as $-(m\omega^2 r+kr^3)$, so that one can explore the region $\Omega \geq \omega$.\\
Condensates rotating in this kind of trap present a very rich variety of vortex phases. At sufficiently slow rotation speeds, the system is not expected to have any vortex. For increasing $\Omega$ there is a sequence of states with more and more vorticity, where singly and multiply quantized vortices may be present at the same time \cite{JaKa,JaKaLu,Lu}.\\
At rotation speeds $\Omega \sim \omega$, a large triangular array of singly quantized vortices appears, which is not qualitatively different from what happens in a purely harmonic trap. The difference is when $\Omega$ is taken larger than $\omega$: the condensate continues to expand radially and there is a critical speed $\Omega_c$ at which the centrifugal force creates a hole in the condensate at the center of the trap \cite{FJS,KB,ABD}. For rotation speeds $\Omega \gtrsim \Omega _c$ there is a vortex lattice in an annular region and a central hole where the density vanishes and around which there is a circulation. However, this hole is not usually referred to as a giant vortex as not all the circulation is contained in a central vortex \cite{FJS}.\\
Finally, for even larger rotation speeds the condensate continues to expand radially with constant area so that the width of the annulus decreases and all the vortices are expected to retreat into the central hole, resulting in a pure irrotational state with macroscopic circulation \cite{FJS,KB,KTU}, often called a giant vortex \cite{AD,D}.\\
In this paper we address the regime $\Omega \gtrsim \Omega_c$. We would like to know the precise repartition of vortices, both in the annular Thomas-Fermi region where we expect that they form a regular lattice, and in the region of low density. Indeed, we believe that there should be "invisible" vortices beyond the external radius of the condensate (as is the case for a harmonically trapped condensate in fast rotation \cite{ABD}) and in the central hole \cite{FJS}.\\
In the limit of weak anharmonicity, the single particle states are restricted to the Lowest Landau Level (LLL) wave functions \cite{ABD}. Thus our analysis consists in minimizing the Gross-Pitaevskii energy in the LLL. Using the explicit expression of the projection onto the LLL we get theoretical results in the spirit of \cite{ABNphi,ABNmath}. We prove that a wave function minimizing the Gross-Pitaevskii energy in the LLL cannot have a finite number of zeroes, thus the vortices of the condensate cannot lie in a bounded domain. This implies that there should be an infinite number of invisible vortices. We also construct critical points for the Gross-Pitaevskii energy by distorting a regular hexagonal vortex lattice in the following way: first, using a theta function as is done for the Abrikosov problem \cite{Abri}, we construct a function $u$ which modulus is periodic over an infinite regular hexagonal lattice and which vanishes only at the points of the lattice with simple zeroes. Next we multiply $u$ by a slow varying profile $\alpha$ (which corresponds to the square root of the coarse-grain average of the atom density over several cells \cite{Ho,ABD}). We then project $\alpha u$ onto the LLL. It can be shown \cite{ABNmath} that projecting this function onto the LLL implies a distortion of the vortex lattice.\\
The state of a condensate in the LLL is entirely known from the positions of its vortices. This allows us to adapt the numerical method of \cite{ABD} and minimize the Gross-Pitaevskii energy functional of the condensate by varying the locations of its vortices. As is expected in the range of parameters we chose, the condensate lies in an annular domain in which vortices are regularly distributed \cite{FJS,KB}. We also find that there are invisible vortices both beyond the outer radius of the condensate and in the central hole. The vortex lattice is strongly distorted at the inner and outer edges of the condensate and we find that multiply quantized vortices appear at the center of the trap.\\
The paper is organized as follows: in Section \ref{sec:Model} we describe our model and general approach. We show how reducing the minimization of the Gross-Pitaevskii energy to the LLL leads to a new variational problem for which we derive an Euler-Lagrange equation satisfied by the minimizer. We then use this equation in Section \ref{sec:theo} to show that the minimizer necessarily has an infinite number of zeroes and to construct critical points for the energy. In Section \ref{sec:num} we describe our numerical method and comment our results, relating them to our theoretical approach. Finally we give our conclusions in Section \ref{sec:Concl}.\\

\section{Model-Approach}\label{sec:Model}

\subsection{Gross-Pitaevskii energy and LLL reduction}\label{sec:Model1}

We assume a strong confinement along the rotation axis, so that we are in an almost 2-D situation. We take $\omega$, $\hbar \omega$ and $\sqrt{\hbar/(m\omega)}$ as units of frequency, energy and length respectively and consider the Gross-Pitaevskii energy of the condensate in the rotating frame, 
\begin{multline}\label{eq:EGP} 
\E _{GP}(\psi)=\int_{\R^2} \left( \frac{1}{2}\left| \nabla \psi -i\Omega x^{\perp}\psi \right|^2 \right. \\ 
\left. +\left( \frac{1-\Omega^2}{2} \vert x\vert ^2 +\frac{k}{4}\vert x\vert ^4 \right) \vert \psi \vert^2 +\frac{G}{2}\vert \psi \vert ^4 \right) dx,
\end{multline}
where $x=(x_1,x_2)$, $x^{\perp}=(-x_2,x_1)$ and $G$ is a dimensionless coefficient characterizing the strength of atomic interactions (see \cite{FJS,ABD} for example). We minimize $\E_{GP}$ under the mass constraint $\int \vert \psi \vert ^2=1$. The main idea of our analysis is to restrict the minimization of $\E_{GP}$ to the first eigenspace of the operator $-\left(\nabla -i\Omega x^{\perp}\right)^2$, corresponding to the eigenvalue $\Omega$. This is the Lowest Landau Level, introduced in the context of Bose-Einstein condensation by Ho \cite{Ho} and then successfully used in different studies on harmonically trapped condensates rotating at speeds close to the trap frequency (see for example \cite{CKR,WBP,ABD}). In \cite{JaKa,JaKaLu} the LLL approximation is used to determine the phase diagram of a condensate trapped by a harmonic plus quartic potential in the limit of weak interactions and small anharmonicity.\\
The LLL consists of functions of the form $\psi(z)=f(z)e^{-\Omega \vert z \vert^2 /2}$ where $z$ is the complex variable $x_1+ix_2$ and $f$ is a holomorphic function. For an LLL function $\psi$, the Gross-Pitaevskii energy reduces to
\begin{multline}\label{eq:ELLL}
\E _{GP}(\psi)=\Omega +\E_{LLL}(\psi) \\
\E_{LLL}(\psi)= \int_{\C} \left( \left(\frac{1-\Omega^2}{2}\vert z \vert ^2  + \frac{k}{4} \vert z \vert ^4 \right)\vert \psi \vert ^2 + \frac{G}{2} \vert \psi \vert ^4\right)dz, 
\end{multline}
so that we will minimize the energy $\E _{LLL}$ in the LLL under the mass constraint $\int \vert \psi \vert ^2=1$.\\
Minimizing (\ref{eq:ELLL}) without the assumption that $\psi$ is in the LLL gives the main scales of the problem. One gets the Thomas-Fermi distribution
\begin{equation}\label{eq:TF1}
\vert \psi_{TF} (r) \vert^2 = \max\left( \frac{\mu+\frac{(\Omega^2 -1)}{2}r^2-
\frac{k}{4}r^4}{G}, 0\right)
 \end{equation}
where $r$ is the radial coordinate. We refer to \cite{FJS,ABD} for the detailed study of such a distribution. In particular one gets the expression of $\Omega _c$ (see equation (\ref{eq:Omegac})), which allows to deduce that for $\vert 1-\Omega \vert$ at most of the order of $k^{2/3}G^{1/3}$ we have $\Omega \sim \Omega_c$. Then $\E_{LLL}(\psi)$ has the order of $k^{1/3}G^{2/3}$ and the spatial extension of $\psi_{TF}$ is of the order of $(G/k)^{1/6}$.\\
The LLL approximation is valid when two conditions are fulfilled \cite{Ho,ABD}: $(i)$ the excess energy $\E_{LLL}(\psi_{TF})$ is small compared to the splitting between the LLL and the first excited Landau level ($\sim 2$ in our units), and $(ii)$ the spatial extension of the condensate is large compared to the vortices interdistance, so that coarse-grain averages of atomic and vortex densities over several vortex cells make sense. Using the scalings derived from (\ref{eq:TF1}), this conditions reduce to 
\begin{equation}\label{eq:reducLLL}
k \ll G \ll \frac{1}{\sqrt{k}} \mbox{ and } \vert 1-\Omega \vert \lesssim k^{2/3}G^{1/3}.
\end{equation}
In Section \ref{sec:theo3} we will give more details on how the scalings (\ref{eq:reducLLL}) are derived and see that taking into account the LLL constraint does not modify the scales of the problem. Indeed, we show in Section \ref{sec:theo2} that although a function such as (\ref{eq:TF1}) cannot be in the LLL, a proper distribution of vortices allows one to approximate very closely such a profile in the LLL, modifying only slightly the averaged atomic density of the condensate.

\subsection{Mathematical framework for the small anharmonicity regime}\label{sec:Model2}
 
We now describe the mathematical framework we are going to use in Section \ref{sec:theo}. It was introduced in \cite{ABNphi,ABNmath} for the theoretical study of an harmonically trapped condensate.\\
We define a small parameter 
\begin{equation}\label{eq:epsilon}
 \varepsilon=k^{1/3}
\end{equation}
corresponding to our small anharmonicity regime and study the asymptotics of the problem as $\varepsilon \rightarrow 0$. We take 
\begin{equation}\label{eq:Omega}
\Omega = 1+\beta k^{2/3} 
\end{equation}
with $\beta \sim G^{1/3}$ so that we are in the range of parameters (\ref{eq:reducLLL}) where the LLL approximation is justified. We rescale distances by making the change of variables $\phi(z)=\sqrt{\varepsilon}\psi(\sqrt{\varepsilon }z)$, the energy becomes $\E _{LLL}(\psi)=\varepsilon E_{LLL}(\phi)$ with
\begin{equation}\label{eq:ELLL2}
E_{LLL}(\phi)= \int_{\C} \left( \left( -\beta \vert z \vert ^2  + \frac{1}{4} \vert z \vert ^4 \right)\vert \phi \vert ^2 + \frac{G}{2} \vert \phi \vert ^4\right)dz.  
\end{equation}
We minimize this new energy amongst all functions $\phi$ in the LLL satisfying the mass constraint $\int \vert \phi \vert^2 =1$.\\
For every $\phi$ in the LLL we define the function $f(z)=\phi(z)e^{\vert z \vert^2/2\varepsilon}$. By definition of the LLL, $f$ belongs to the Fock-Bargmann space
\begin{equation}\label{eq:Fepsilon}
\F _{\varepsilon}=\left\lbrace f \mbox{ holomorphic }, \int_{\C} \vert f\vert^2 e^{-\vert z \vert^2/\varepsilon}dz < \infty \right\rbrace. 
\end{equation}
The space $\F_{\varepsilon}$ is a Hilbert space for the scalar product 
\begin{equation*}
\left\langle f,g \right\rangle=\int_{\C} \overline{f(z)}g(z) e^{-\vert z \vert^2/\varepsilon}dz.
\end{equation*}
The point of introducing such a space is that the orthogonal projection of any function $g$ onto $\F_{\varepsilon}$ is explicitly known \cite{Mar,Fol}:
\begin{equation}\label{eq:Piepsilon}
\Pi_{\varepsilon}(g)(z)=\frac{1}{\pi \varepsilon}\int_{\C}e^{z\bar{z'}/\varepsilon}e^{-\vert z' \vert^2 /\varepsilon} g(z')dz',
\end{equation}
so that if we write the variational problem (\ref{eq:ELLL2}) for $f$ rather than for $\phi$, namely if we look for $f\in \F_{\varepsilon}$ minimizing
\begin{multline}\label{eq:GLLL}
F_{LLL}(f)=\int_{\C} \left(\left( -\beta \vert z \vert ^2  + \frac{1}{4} \vert z \vert ^4 \right)\vert f \vert ^2 e^{-\vert z \vert^2 /\varepsilon} \right.\\ 
\left. + \frac{G}{2} \vert f \vert ^4 e^{-2\vert z \vert^2 /\varepsilon}\right)dz
\end{multline}
 under the constraint $\left\langle f,f\right\rangle=1$, we are able to derive that $f$ satisfies the equation
\begin{equation}\label{eq:EEL1}
-\beta M_{\varepsilon} f+\frac{1}{4}\left( M_{\varepsilon}^2 f + \varepsilon M_{\varepsilon} f\right)+G\Pi_{\varepsilon}(e^{-\vert z \vert^2 /\varepsilon}\vert f\vert^2 f)=\mu f 
\end{equation}
where $\mu$ is the chemical potential coming from the mass constraint and $M_{\varepsilon}$ is the operator defined by
\begin{equation*}
 M_{\varepsilon}= \varepsilon \partial_z z .
\end{equation*}
Indeed, the weak derivative of the energy $F_{LLL}$ at $f$ along $g$ is given by
\begin{multline*}
DF_{LLL}(f) \cdot g = \int_{\C} \left(\left( -\beta \vert z \vert ^2  + \frac{1}{4} \vert z \vert ^4 \right)\bar{f}g e^{-\vert z \vert^2 /\varepsilon} \right. \\ 
\left. + \frac{G}{2} \vert f \vert ^2\bar{f}g e^{-2\vert z \vert^2 /\varepsilon}\right)dz.
\end{multline*}
Using integration by parts and $\partial_{\bar{z}}f=\partial_{\bar{z}}g=0$ on the first term and the fact that $\Pi_{\varepsilon}(g)=g$ on the second term, we obtain (\ref{eq:EEL1}).\\
We give two equivalent forms of (\ref{eq:EEL1}) that we shall need in the sequel:
\begin{equation}\label{eq:EEL2}
-\beta M_{\varepsilon} f+\frac{1}{4}\left( M_{\varepsilon}^2 f + \varepsilon M_{\varepsilon} f\right)+\frac{G}{2}\bar{f}(\varepsilon \partial_{z})[f^2(z/2)]=\mu f, 
\end{equation}
\begin{multline}\label{eq:EEL3}
(-\beta+\frac{1}{4}\varepsilon)\Pi_{\varepsilon}(\vert z \vert^2 f)+\frac{1}{4}\Pi_{\varepsilon}\left( \vert z\vert^2 \Pi_{\varepsilon}(\vert z\vert^2 f)\right) \\ 
 +G\Pi_{\varepsilon}(^{-\vert z \vert^2 /\varepsilon}\vert f\vert^2 f)=\mu f.
\end{multline}
The operator $\bar{f}(\varepsilon \partial_{z})$ is defined by 
\[
 \bar{f}(\varepsilon \partial_{z})[g]=\sum ^{+\infty}_{k=0}\overline{a_k}(\ep \partial _z)^k g
\]
if $f(z)=\sum a_k z^k$. Equation (\ref{eq:EEL2}) is obtained from (\ref{eq:EEL1}) as in \cite{ABNphi} with some algebra on the non-linear term. To get (\ref{eq:EEL3}) we use an integration by parts to show that $M_{\varepsilon}f=\Pi_{\varepsilon}(\vert z\vert^2 f)$, then $M_{\varepsilon}^2f=\Pi_{\varepsilon}(\vert z\vert^2 \Pi_{\varepsilon}(\vert z\vert^2 f))$ and we get the result.\\
In the following section we prove our theoretical results. Using (\ref{eq:EEL2}) we show below that any minimizer $f$ of (\ref{eq:GLLL}) (and thus any minimizer $\phi$ of (\ref{eq:ELLL2})) has an infinite number of zeroes for $\varepsilon$ small enough. The construction of critical points is done using (\ref{eq:EEL3}).\\

\section{Analytical study}\label{sec:theo}

In this section we present the results we are able to derive from equations (\ref{eq:EEL2}) and (\ref{eq:EEL3}).

\subsection{Infinite number of zeroes}\label{sec:theo1}

We show that any minimizer $f$ of (\ref{eq:GLLL}) (and therefore any minimizer $\psi$ of (\ref{eq:ELLL})) has an infinite number of zeroes if $\ep$ is small enough. The argument is by contradiction and in two steps.
\begin{enumerate}
\item Suppose $f$ has a finite number of zeroes. Then one may write $f(z)=P(z)e^{\varphi(z)}$ where $P$ is a polynomial and $\varphi$ is a holomorphic function. Now $f\in \F_{\ep}$ and the condition $\int_{\C} \vert f\vert^2 e^{-\vert z \vert^2/\varepsilon}dz < \infty$ implies that $Re(\varphi(z))\leq \vert z\vert^2 /(2 \ep)$. It is well-known (see \cite{Boa} for example) that a holomorphic function can satisfy this condition only if it is a polynomial of degree less than 2. Therefore we know that 
\begin{equation}\label{eq:f=Pephi}
 f(z)=P(z)e^{\alpha_1 z +\alpha_2 z^2}
\end{equation}
and the integrability condition on $f$ implies $\alpha_2 \leq 1/(2 \ep)$. Injecting (\ref{eq:f=Pephi}) in (\ref{eq:EEL1}) and comparing the exponential growth of the different terms of (\ref{eq:EEL1}) as in \cite{ABNmath} yields $\alpha_1=\alpha_2=0$. So, if $f$ has a finite number of zeroes, it is a polynomial.
\item Now, suppose $f$ is a polynomial of degree $n$ and inject this in (\ref{eq:EEL2}). The term $-\beta M_{\varepsilon} f+\frac{1}{4}\left( M_{\varepsilon}^2 f + \varepsilon M_{\varepsilon} f\right)$ is a polynomial of degree $n$, therefore (\ref{eq:EEL2}) implies that the term $\frac{G}{2}\bar{f}(\varepsilon \partial_{z})[f^2(z/2)]$ is also of degree $n$. But $(\ep \partial _z )^k[f^2(z/2)]$ is of degree $2n-k$, so that $f$ must be of the form $f(z)=cz^n$. Injecting this a last time in (\ref{eq:EEL2}), using the improved Stirling \cite{Rob} formula and the condition $\left\langle f,f \right\rangle=1$ yields a condition on $n$:
\begin{multline}\label{eq:n}
\mu + \beta \ep \geq -\beta n \ep + \frac{Ge^{-1/12}}{2\pi \ep \sqrt{n}} \\ 
+\frac{\ep^2}{2}+\frac{n^2 \ep^2}{4}+\frac{3n\ep^2}{4}.
\end{multline}
We now bound the chemical potential $\mu$: taking the $\F_{\ep}$-scalar product of each side of (\ref{eq:EEL1}) with $f$ yields
\begin{equation}\label{eq:bornemu}
 \mu \leq 2F_{LLL}(f) +\beta^2 - \frac{\beta \ep}{2}+\frac{\ep^2}{16}.
\end{equation}
Here we used the fact that the spectrum of $-\beta M_{\varepsilon} +\frac{1}{4}\left( M_{\varepsilon}^2  + \varepsilon M_{\varepsilon} \right)$ is bounded below uniformly with respect to $\ep$. We shall see in Section \ref{sec:theo3} (see (\ref{eq:Energie})) that 
\begin{multline*}
F_{LLL}(f)=E_{LLL}(fe^{-\vert z \vert^2/2\ep}) \\ 
\leq \ep ^{-1}\E_{LLL}=\left(\frac{3}{5}\left(\frac{3bG}{8\pi}\right)^{2/3}-\beta ^2\right).
\end{multline*}
Thus we have
\begin{multline}\label{eq:n2}
2\ep ^{-1}\E_{LLL}+\beta^2  - \frac{\beta \ep}{2}-\frac{7\ep^2}{16} \\ 
\geq -\beta n \ep + \frac{Ge^{-1/12}}{2\pi \ep \sqrt{n}}+\frac{n^2 \ep^2}{4}+\frac{3n\ep^2}{4}
\end{multline}
and the left-hand side of (\ref{eq:n2}) is bounded uniformly with respect to $\ep$. Minimizing the right-hand side of (\ref{eq:n2}) with respect to $n$ (taken as a continuous variable as it should be very large when $\ep$ is small) for fixed $\ep$ yields
\begin{equation}\label{eq:n3}
n \sim \left( \frac{Ge^{-1/12}}{2\pi} \right) ^{2/5}\ep ^{-6/5}
\end{equation}
and
\begin{multline}\label{eq:borneep}
\frac{6}{5}\left(\frac{3bG}{8\pi}\right)^{2/3}-\beta ^2 - \frac{\beta \ep}{2}-\frac{7\ep^2}{16} \\ \geq \frac{5}{4} \left( \frac{Ge^{-1/12}}{2\pi} \right)^{4/5} \ep^{-2/5}+O(\ep ^{-1/5})
\end{multline} 
which is a contradiction if $\ep$ is small enough.\\
\end{enumerate}
In the rest of the paper we take an $\ep$ small enough for the inequation (\ref{eq:borneep}) not to be verified. We then conclude that $f$ has an infinite number of zeroes. As there is a limit on how close vortices can be in this regime (see \cite{Ho}), they cannot lie in a bounded domain and the vortex pattern extends to infinity.

\subsection{Construction of critical points}\label{sec:theo2}

As is the case for harmonically trapped condensates in fast rotation, we expect that in the range of parameters we explore the scales of the problem will decouple. Namely we expect any minimizer of (\ref{eq:ELLL2}) to be of the form $\alpha u$ where $u$ varies on the scale of the vortex pattern (which is small compared to the size of the condensate) and $\alpha$ is a slow varying profile giving the general shape of the condensate. In this section we show that although such a function is not in the LLL, one can approach it by an LLL function which is an almost critical point for the energy (\ref{eq:ELLL2}). We also relate the effect of the projection onto the LLL to the distortion of a regular lattice in the region of low atomic density.\\
More precisely we introduce 
\begin{equation}\label{eq:Abriko}
  u _{\tau}(z)=e^{-{|z|^{2}/2\ep}}f _{\tau}(z),\quad
  f _{\tau}(z)=e^{ z^{2}/2\ep}\Theta\left(\sqrt{\frac{\tau_{I}}{\pi \ep}}z, \tau\right)
\end{equation}
where $\tau=\tau_R+i\tau_I$ is any complex number and 
\begin{equation}\label{eq:theta}
\Theta(v,\tau)=\frac{1}{i}\sum_{n=-\infty}^{+\infty}(-1)^{n}e^{i\pi\tau(n+1/2)^{2}}
e^{(2n+1)\pi iv}. 
\end{equation} 
The $\Theta$ function has the property (see \cite{Cha} for more details) $\Theta(v+k+l\tau,\tau)=(-1)^{k+l}e^{-2i\pi lv} e^{-i\pi l\tau}\Theta(v,\tau)$ so that $|u _{\tau}(z)|$ is periodic over the lattice $\sqrt{\frac{\pi\ep}{\tau_I}}\Z\oplus \sqrt{\frac{\pi\ep}{\tau_I}}\Z\tau$, and vanishes at each point of the lattice.\\
The interest of introducing such functions is twofold. Firstly it is known \cite{Cha} that any function $v$ whose modulus is periodic over the lattice $\sqrt{\frac{\pi\ep}{\tau_I}}\Z\oplus \sqrt{\frac{\pi\ep}{\tau_I}}\Z\tau$, vanishes exactly on the points of the lattice with simple zeroes and such that $g=ve^{\vert z \vert^2/2\ep }$ is holomorphic must be proportional to $u_{\tau}$. Secondly, the function $f_{\tau}$ is a solution to the Abrikosov problem (see \cite{Abri,Tor,ABNphi,ABNmath})
\begin{equation}\label{eq:fabri}
\Pi(|f_{\tau} |^2e^{-|z|^2/\ep}f_{\tau} )=\lambda_{\tau} f_{\tau} ,\mbox{ with } \lambda_{\tau}=\left< |u _{\tau} |^2\right> b(\tau),
\end{equation} 
and 
\begin{equation}\label{btau}
b(\tau)=\frac {\left< |u _{\tau}|^4\right>}{\left < |u _{\tau} |^2\right>^2}=\sum_{k,l\in \Z} e^{-\pi |k\tau -l|^2/\tau_I}.
\end{equation}
Equation (\ref{eq:fabri}) is similar to (\ref{eq:EEL3}) without the potential term and with $\mu=\lambda_{\tau}$ so that one can expect to obtain a solution of (\ref{eq:EEL3}) by a slight modification of $f_{\tau}$. We refer to \cite{ABNphi} and the references therein for details on the quantity $b(\tau)$. Let us just mention that it is minimum ($b(\tau)\sim 1.16$) for $\tau = e^{2i\pi / 3}$, which corresponds to a hexagonal lattice.\\
We define 
\begin{equation}\label{eq:deff}
f_{\alpha,\tau}=\frac{\Pi_{\ep}(\alpha f_{\tau})}{\left\langle \Pi_{\ep}(\alpha f_{\tau}),\Pi_{\ep}(\alpha f_{\tau}) \right\rangle^{1/2}}  
\end{equation}
and
\begin{equation}\label{eq:defu}
 u_{\alpha,\tau}=f_{\alpha,\tau}e^{-\vert z \vert^2 / 2\ep}
\end{equation}
where $\alpha \in C^{0,1/2}(\C,\C)$ is a slow varying profile with compact support and $\int \vert \alpha \vert ^2 =1$. We use Lemma 5.4 in \cite{ABNmath}, which states that 
\begin{equation}\label{eq:ualpha}
u_{\alpha,\tau} = \alpha u_{\tau}+O(\ep^{1/4}).
\end{equation}
This, together with the fact that $| u_{\tau} |^2$ is periodic over a lattice of period bounded by $O(\ep ^{1/2})$ yields
\begin{multline}\label{eq:resultEq}
(-\beta+\frac{1}{4}\varepsilon)\Pi_{\varepsilon}(\vert z \vert^2 f_{\alpha,\tau})+\frac{1}{4}\Pi_{\varepsilon}\left( \vert z\vert^2 \Pi_{\varepsilon}(\vert z\vert^2 f_{\alpha,\tau})\right)\\ 
+G\Pi_{\varepsilon}(e^{-\vert z \vert^2 /\varepsilon}\vert f_{\alpha,\tau}\vert^2 f_{\alpha,\tau})-\mu f_{\alpha,\tau}= \\
\Pi_{\ep}((-\mu -\beta | z |^2 +\frac{1}{4}| z |^4+Gb(\tau)| \alpha  |^2)\alpha \Pi_{\ep}f_{\tau})+O(\ep^{1/4})
\end{multline}
 and 
\begin{multline}\label{eq:resulEn}
F_{LLL}(f_{\alpha,\tau})=E_{LLL}(u_{\alpha,\tau})= \\
\int_{\C} \left( \left( -\beta \vert z \vert ^2  + \frac{1}{4} \vert z \vert ^4 \right)\vert \alpha \vert ^2 + \frac{b(\tau)G}{2} \vert \alpha \vert ^4\right)dz +O(\ep ^{1/4}).
\end{multline}
Equations (\ref{eq:resultEq}) and (\ref{eq:resulEn}) are the scale decoupling we expected: the only contribution of the lattice to the energy is through the coefficient $b(\tau)$, which is minimum for the hexagonal lattice, $\tau = e^{2i\pi /3}$. From now on we shall note $b=b(e^{2i\pi /3})\sim 1.16$. We now have to minimize (\ref{eq:resulEn}), which is an energy similar to (\ref{eq:ELLL}), with respect to the profile $\alpha$ which is not in the LLL, so that the minimization is performed in the usual way. We obtain the Thomas-Fermi profile
\begin{equation}\label{eq:alpha}
 |\alpha|^2 (z) = \max\left( \frac{\nu +\beta |z|^2-\frac{1}{4}|z|^4}{bG}, 0\right)
\end{equation}
where $\nu$ is the chemical potential associated with the constraint $\int |\alpha|^2=1$ and $b$ takes into account the vortices contribution. With such a profile and taking $\mu=\nu$ in (\ref{eq:resultEq}) we get 
\begin{multline}\label{eq:resultEq2}
(-\beta+\frac{1}{4}\varepsilon)\Pi_{\varepsilon}(\vert z \vert^2 f_{\alpha,\tau})+\frac{1}{4}\Pi_{\varepsilon}\left( \vert z\vert^2 \Pi_{\varepsilon}(\vert z\vert^2 f_{\alpha,\tau})\right)\\ 
+G\Pi_{\varepsilon}(e^{-\vert z \vert^2 /\varepsilon}\vert f_{\alpha,\tau}\vert^2 f_{\alpha,\tau})=
\nu f_{\alpha,\tau}+O(\ep^{1/4}) 
\end{multline}
so that $f_{\alpha,\tau}$ is almost a critical point for the energy (\ref{eq:GLLL}).\\
We now explain the effect of our procedure on the vortex pattern. The property $u_{\alpha,\tau}=\alpha u_{\tau} + O(\ep ^{1/4})$ shows that on the support of $\alpha$ the zeroes of $u_{\alpha,\tau}$ are distributed close to those of $u_{\tau}$, on a regular hexagonal lattice. To get information on the "invisible vortices" lying in the region of low density, we evaluate the number $N(R)$ of zeroes of $f_{\alpha,\tau}$ in a ball of radius $R$. The Cauchy formula yields 
\begin{equation}\label{eq:Cauchy}
N(R)=\frac{R}{2\pi \ep}\int_0^{2\pi}\!\!\!\! d\theta \frac{\int e^{R\overline{z'}/\ep}e^{-|z'|^{2}/\ep}\overline{z'}\alpha(z')
f_{\tau}(z'e^{-i\theta})dz'}{\int e^{R\overline{z'}/\ep}e^{-|z'|^{2}/\ep}\alpha(z')
 f_{\tau}(z'e^{-i\theta})dz'}.
\end{equation} 
Using a Laplace method to evaluate the integrals, we see that their ratio is bounded for large $R$, so that $N(R)\propto R/\ep$. A regular lattice would give $N(R)\propto R^2/\ep$, so we deduce that the lattice is strongly distorted outside the external radius of the condensate. Note that this method does not give information on the distribution of vortices in the central hole when the condensate has an annular form, for this would correspond to $R$ small, and the Laplace method is not efficient in this case. We refer to our numerical simulations in Section \ref{sec:num} for the vortex structure in the central hole.

\subsection{Evaluation of physical quantities}\label{sec:theo3}

Here we analyze the TF profile (\ref{eq:alpha}), or rather $\tilde{\alpha}(z)=\sqrt{\ep}^{-1}\alpha(z\sqrt{\ep}^{-1})$, to evaluate some relevant physical quantities in the original scaling.\\
The analysis of a profile such as (\ref{eq:alpha}) has already been done in \cite{FJS,ABD}, so we only adapt and summarize the results. The critical rotation speed for the condensate to develop a central hole is 
\begin{equation}\label{eq:Omegac}
\Omega _c = 1+\left(\frac{3k^2bG}{8\pi}\right)^{1/3}.
\end{equation}
For subcritical velocities, the behavior of the condensate is not qualitatively different from that of a harmonically trapped condensate, so we focus on velocities $\Omega \geq \Omega_{c}$ which is equivalent to $\beta \geq \left(\frac{3bG}{8\pi}\right)^{1/3}$. Then the inner and outer radius of the condensate $R_{\pm}$ are given by the relations
\begin{equation}\label{eq:Rayons}
R_+ ^2 + R_- ^2 = 4\beta k^{-1/3}, \quad R_+ ^2 - R_- ^2 = \left(\frac{24bG}{k}\right) ^{1/3},
\end{equation}
the chemical potential is
\begin{equation}\label{eq:mu}
\mu=\left(\left(\frac{3bG}{8\pi}\right)^{2/3}-\beta^{2}\right)k^{1/3}
\end{equation}
and the energy is  
\begin{equation}\label{eq:Energie}
\E_{LLL}=\left(\frac{3}{5}\left(\frac{3bG}{8\pi}\right)^{2/3}-\beta ^2\right)k^{1/3}.
\end{equation}
This is slightly different from the result obtained in \cite{FJS} with the solid-body approximation. Indeed, our analysis allows one, through the coefficient $b$, to take into account more precisely the contribution of the vortex patterns to the energy and atomic density, whereas the solid-body approximation implies an infinite regular lattice with prescribed volume of the cell. These results and the fact that $b\sim 1.16$ imply that our analysis is valid within the range of parameters we announced in Section 2.1 : equations (\ref{eq:Rayons}) yield that the spatial extension of the condensate is large compared to $1$ (which is the order of the vortex pattern spacing) if 
\[
 k\ll G.
\]
We now see that the condition $\beta \sim G^{1/3}$ (or $\vert 1-\Omega \vert \lesssim k^{2/3}G^{1/3}$) is equivalent to $\Omega \sim \Omega_c$ and ensures that the energy $\E_{LLL}$ is of order $k^{1/3}G^{2/3}$, so that the energy condition for the LLL approximation to be valid reduces to
\[
 G\ll \frac{1}{\sqrt{k}}.
\]
Now, as we have both $k\ll G$ and $G\ll \frac{1}{\sqrt{k}}$, we see that necessarily $k\ll 1$, which justifies our study of a small anharmonicity regime.
 
\section{Numerical Simulations}\label{sec:num}

\subsection{Approach}\label{sec:num1}

We want to numerically approach the minimizer $\phi$ of $\E_{LLL}$ in the LLL. We write 
\begin{equation}\label{eq:LLLnum}
 \phi(z)=P(z)e^{-\Omega \vert z \vert ^2 /2}
\end{equation}
with $P$ a holomorphic function. As polynomials are dense in $\F_{\ep}$, it is reasonable to fix an integer $n$ and to restrict the analysis to functions $\phi$ where $P$ is a polynomial of degree less than $n$. We write our trial functions as in \cite{ABD}, where $A=\Vert \phi \Vert _{L^2}^{-1/2} $ is the normalization factor:
\begin{equation}\label{eq:ftest}
\phi(z)=A \prod ^n _{j=1} (z-z_j)e^{-\Omega \vert z \vert ^2 /2}
\end{equation}
and vary the locations $z_j$ of vortices. An alternative method (used for example in \cite{CKR} for a harmonically trapped condensate) would be to take 
\begin{equation*}
\phi(z)=A\left( \sum ^n _{j=0} b_j z^j \right) e^{-\Omega \vert z \vert ^2 /2}
\end{equation*}
and vary the coefficients $b_j$. The interest of our approach is to give a direct access to the exact repartition of vortices, whereas the alternative method would require to compute the roots of a polynomial of degree $n$, which is a delicate task for large $n$. In particular, varying the coefficients could probably not give the precise locations of invisible vortices.\\
We used a conjugate gradient method with a Goldstein and Price line-search. The integrals are computed using the Gauss-Hermite method, and we take enough Gauss points for the computations to be exact. This results in quite expensive calculations, but we have been able to numerically construct condensates with up to $\sim 120$ vortices.

\subsection{Results}\label{sec:num2}

\begin{figure}[h]
\begin{centering}
\includegraphics[width=85mm]{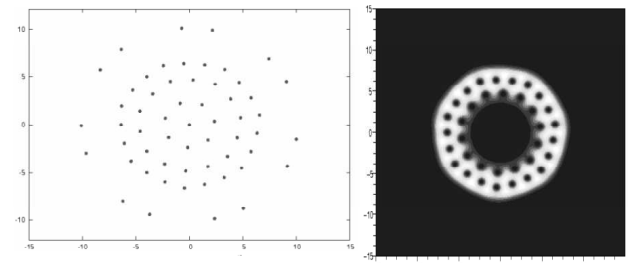}
\caption{Vortex structure and atomic density for $G=3, k=10^{-4}, \beta=1 \: (\Omega-1 = 2.2 \, 10^{-3})$. There are 67 vortices in total, the central vortex is constituted of 11 single vortices. \label{fig1}}
\end{centering}
\end{figure}
\begin{figure}[h]
\begin{centering}
\includegraphics[width=85mm]{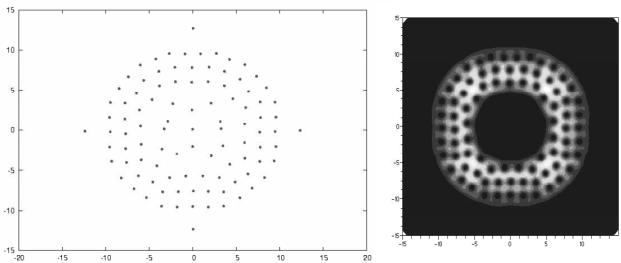}
\caption{Vortex structure and atomic density for $G=3, k=10^{-5}, \beta=1 \: (\Omega-1 = 4.6 \, 10^{-4})$. There are 119 vortices in total, the central vortex is constituted of 20 single vortices. \label{fig2}}
\end{centering}
\end{figure}

We show in Fig. \ref{fig1} and Fig. \ref{fig2} typical examples of configurations we numerically computed. The qualitative features of the vortex patterns and atomic densities confirm our theoretical results and are in good agreement with existing theoretical and numerical studies \cite{FJS,KB}. Note however that the numerics become quite intricate for large number of vortices, which accounts for the relative lack of symmetry of Fig. \ref{fig2}.\\
\begin{figure}[h]
\begin{centering}
\includegraphics[width=85mm]{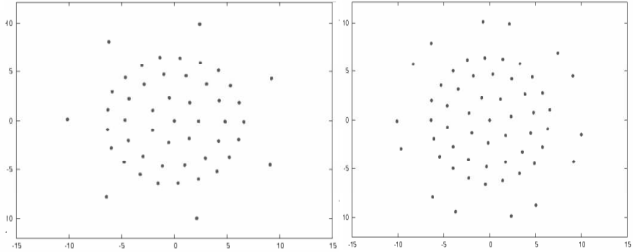}
\caption{Two example of vortex configurations for $G=3, k=10^{-4}, \beta=1 \: (\Omega-1 = 2.2 \, 10^{-3})$, respectively with $n=60$ and $n=67$ vortices. \label{fig3}}
\end{centering}
\end{figure}
As was expected, the condensate develops a central hole and visible vortices are regularly distributed in the annular region of significant atomic density. Our computations also show a distortion of the vortex pattern near the external radius of the condensate, resulting in invisible vortices in the exterior region of low atomic density as is the case for harmonically trapped condensates \cite{ABD}. Some vortices also lie in the central hole as theoretically predicted (see for example \cite{FJS}) and we can get information on their precise locations: we observe a distortion of the regular lattice near the inner radius of the condensate, resulting in isolated singly-quantized vortices encircling a central multiply quantized vortex. We computed configurations for which this central vortex has up to $20$ units of circulation while there is a total of $32$ units of circulation in the entire hole and $83$ visible vortices. The number of vortices (both visible and invisible) increases with increasing $\Omega$ or $\beta$, but since our scaling does not allow to explore a large domain of $\Omega$ when $G$ is fixed, we mainly varied $k$. The total vorticity of the system increases with decreasing $k$.\\
All vortices do not have the same contribution to the energy: as $n$ increases, the vortex pattern in the annular region of significant atomic density remains the same up to possible rotations, with the additional vortices first gathering in the central vortex, then constituting the distorted lattice near the inner boundary of the condensate and finally occupying the distorted sites beyond the external radius. With increasing $n$ (see Fig. \ref{fig4}), the energy reaches a first plateau when the central figure is formed, constituted of the visible vortices and the vortices in the central hole. A second plateau is reached when enough distorted sites beyond the external radius are occupied. For example, for the parameters corresponding to Fig. \ref{fig1} and Fig. \ref{fig4} ($k=10^{-4},\: G=3, \: \beta=1$), the energy varies by $\sim \pm 10^{-6}$ in relative value when $n$ is increased from $60$ to $67$ (Fig. \ref{fig3}) and the atomic density does not vary significantly. We find good agreement of our numerical results and analytical study, with energies typically differing by $\sim 10^{-3}$ to $\sim 10^{-2}$ in relative value. 

\begin{figure}[ht]
\begin{centering}
\includegraphics[width=85mm]{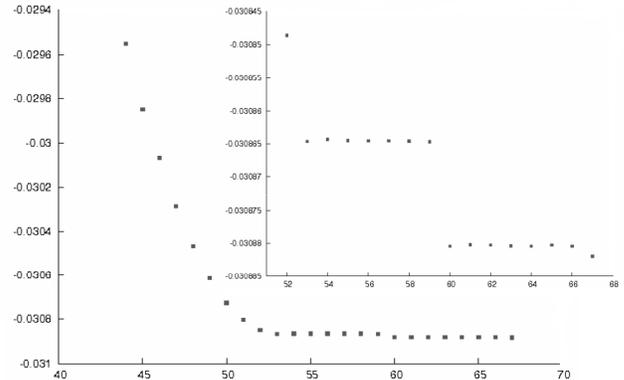}
\caption{Minimum energy as a function of the number of vortices in the trial wave function ($G=3, k=10^{-4}, \beta=1$). \label{fig4}}
\end{centering}
\end{figure}

\section{Conclusion}\label{sec:Concl}
 
We have investigated the vortex structure of a two dimensional annular Bose-Einstein condensate rotating in a quadratic plus quartic confining potential. We focused on the regime where the state of reference is a vortex lattice encircling a central hole carrying a macroscopic circulation. We developed a theoretical and numerical approach based on a small anharmonicity regime and a reduction to the Lowest Landau Level states. Using a theoretical method developed in \cite{ABNphi,ABNmath} we showed that there is an infinite number of vortices in the condensate so that they cannot lie in a bounded domain. We then analytically and numerically constructed critical points for the reduced Gross-Pitaevskii energy and get further information on their vortices. We find that they are regularly distributed on a hexagonal lattice in the annular Thomas-Fermi region where the atomic density takes significant values, and that this pattern is strongly distorted at both edges of the annulus. In particular we find numerical evidence of multiply quantized vortices appearing at the center of the trap. Our results agree with and complete existing studies \cite{FJS,KB,ABD}, showing how the distortion of the vortex pattern modifies the results obtained with the solid-body approximation.

\acknowledgments

 This work was supported by a grant from Région Ile-de-France.

\nocite{*}
\bibliography{QuartiqueLLL}

\end{document}